\documentclass[preprint,showpacs,preprintnumbers,amsmath,amssymb]{revtex4}

\usepackage{graphicx}
\input{epsf}
\begin{document}
\title{Photovoltage in curved 1D systems.}
 \author{M.V. Entin$^{(1)}$, L.I. Magarill$^{(1,2)}$}
\affiliation{$^{(1)}$Institute of Semiconductor Physics, Siberian
Branch of Russian Academy of Sciences, Novosibirsk, 630090, Russia
\\ $^{(2)}$ Novosibirsk State University, Novosibirsk, 630090, Russia}

\begin{abstract}
Curvature of quantum wire results in intrasubband absorption of
 IR radiation that  induces stationary
photovoltage in presence of circular polarization. This effect is
studied in ballistic (collisionless) and kinetic regimes. The
consideration is concentrated on quantum wires with curved central
part. It is shown, that if mean free path is shorter than length
of the curved part the photovoltage does not depend on the wire
shape, but on the total angle of rotation of wire tangent. It is
not  the case when mean free path is finite or large. This
situation was studied for  three specific shapes of wires: ''hard
angle'', ''open book'' and ''$\Omega$-like''.

\end{abstract}


\maketitle

\subsection*{Introduction}
The stationary current induced by alternating force was a subject
of numerous publications.  The ordered motion of electrons in
absence of stationary driving force implies the simultaneous
existence of energy and momentum sources. While the energy
originates from the alternating force itself, the momentum can be
constructively transferred from scattering events via vector
asymmetry pointing the direction of current or from the wave
itself.

The directions of study can be classified in relation to the
participation of light momentum, the source of anisotropy, the
system uniformity, coherence of light etc. In particular, the term
''photogalvanic effect'' \cite{belin,ivch,we} is used to describe
the stationary photocurrent in homogeneous medium with low
symmetry, where the direction of current is determined by the
electric field of light together with the third rank tensor
belonging to the medium itself, while the directional motion is
caused by the participation of electron scattering. The term
''photon drag'' \cite{dan,grinb,gibs} relates to currents due to
the transmission of momentum from photons to electrons and does
not need participation of the ''third body'', namely scatterers.
The term ''quantum pumps'' \cite{tau,brouw,moskal,we1}  is applied
mainly to local quantum systems driven by periodically changing
parameters. The term ''ratchet'' \cite{hanggi,reimann,semidisk1}
is used to describe the stationary flow caused by alternating
force, not necessarily electrical.

The purpose of the present paper is IR photoresponses in curved
quantum wires. The main idea is that in spite of uniformity of IR
electric field the acting component of electric field tangential
to the narrow wire becomes non-uniform with the characteristic
length dictated by the wire curvature. In this sense the situation
reminds non-uniformity of acting component of magnetic field in
the curved 2D systems. We have studied photocurrent in a spiral
quantum wire earlier \cite{spiral}. The system under examination
differs from \cite{spiral} by non-homogeneous curvature. Note,
that another non-uniformly curved system, namely, curved 1D
quantum dot lattice subjected to IR radiation was studied in
\cite{mahm}.

We consider  planar quantum wires with kinks (examples are shown
in the figure) subjected to normally \ incident \ arbitrarily (in
particular, circular) polarized electromagnetic wave. The wire is
assumed to be strictly one-dimensional. In our approach only
tangential component of the electric field acts on electrons in a
classical manner. Owing to curvature the inhomogeneity presents in
some part of the wire. This produces a stationary current in
closed-circuit or a voltage in open-circuit regimes. The problem
is studied in the framework of the classical Boltzmann kinetic
equation for freely moving electrons along the curved quantum wire
(the case of single subband occupied with intrasubband
absorption). The potential (expressed via dependence of lowest
subband on coordinates) caused by inhomogeneity or curvature of
the wire
  will be neglected.

In the lowest order on intensity the stationary current is a
second-order response to electric field ${\bf E}(t)$. In
particular, for coherent monochromatic radiation ${\bf
 E}(t)=\mbox{Re}({\bf E}^\omega e^{-i\omega t})$, ${\bf E}^\omega=E^\omega(1,
i\zeta)/\sqrt{2}$ is the complex amplitude of wave; the degree of
circular polarization $\zeta=0$ for linear polarized and
$\zeta=\pm 1$ for fully circular polarized wave. The considered
shapes of wires can be globally characterized by a vector ${\bf
b}$. (See Fig., where up and down directions are not equivalent
and the vector ${\bf b}$ is vertical.) The direction of stationary
current ${\bf J}$ can be constructed only as a vector product of
${\bf b}$ and the vector product of electric field complex
amplitude ${\bf E}^\omega$ and its complex conjugate: ${\bf
J}\propto [{\bf b}[{\bf E}^\omega{\bf E}^{-\omega}]] $. Thus, the
current should be proportional to $\zeta$. This describes the
phenomenology of the effect: it should exist for circular
polarized radiation and should vanish for linear polarization, and
change sign with the sign of circular polarization. So this effect
can be attributed to the class of circular photogalvanic effects.

 Generally, the value of
current is determined by the wire shape and its symmetry. Unlike
photogalvanic effects (appearing in homogeneous media) the
considered effect has local nature, what makes it related to
quantum pumps. The current is caused by the momentum transfer from
the wire to electrons via inhomogeneity of acting alternating
force. From this point of view the considered phenomenon is some
sort of the photon drag effect.

Both homogeneity of the system  and smallness   of electromagnetic
field wave vector, especially in the case of far IR light,
restrict the known photocurrents. Smallness of the wave vector
leads to relatively weak photon drag effect; the photogalvanic
effect being caused by participation of the scattering by a third
body (impurities, phonons) is also weak. Curved 1D or 2D systems
present possibility to overcome this weakness: the uniform
external field acts on propagating electrons by non-uniform
effective force whose characteristic length, in principle, is
comparable with the Fermi wavelength.

\begin{figure}[h]
\centerline{ \epsfbox{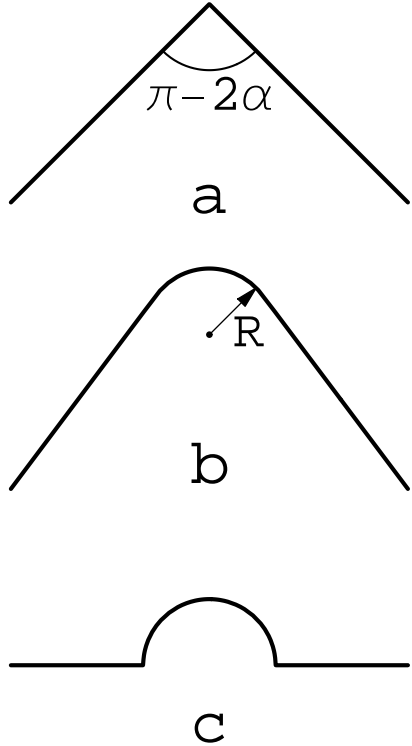} }
 \caption{Considered quantum
wires: a) ''angle'', b) ''open book'' (straight lines tangential
to a circular segment), c) ''$\Omega$-curve'' (semicircle with
straight source and drain). In  cases a) and b) the angle between
straight parts of wire equals $\pi-2\alpha$, radius of circular
segments in cases b) and c) equals $R$.}
\end{figure}

\subsection*{Basic equations}
Let us consider the one-dimensional quantum wire described by a
planar curve ${\bf a}(s)=(a_x,a_y)$ with the tangent ort ${\bf
t}(s)={\bf a}'(s)$, s is the length along the curve. The effective
electric field acting on electrons ${\cal E}$ is determined by the
projection of electric field ${\bf E}(s,t)$ to the tangent ${\cal
E}(s,t)= {\bf E}(t){\bf t}(s)$. We will consider the effect in the
approximation of the classical kinetic equation
\begin{equation}\label{kin}
    \frac{\partial f}{\partial t}+v\frac{\partial f}{\partial s}-e{\cal E}\frac{\partial f}{\partial
    p}=-\frac{f-<f>}{\tau},
\end{equation} where $\tau=1/\nu$ is the relaxation time,
$<f>=(f(p)-f(-p))/2$. We neglect the static potential in the wire,
in particular, connected with the curvature itself. Further we
consider the effect of the monochromatic electric field ${\cal
E}(t)=\mbox{Re}({\cal E}^\omega e^{-i\omega t})$.

We will deal with  the specific cases of symmetric planar curves
depicted in Fig. 1. Corresponding tangent orts are
\begin{eqnarray}\label{angle}
    &&t_x(s)= \cos\alpha,\nonumber\\
    &&t_y(s)=-\mbox{sign}(s)\sin\alpha ~~~~(\mbox{''angle''})
\end{eqnarray}
\begin{eqnarray}\label{openbook}
    &&t_x(s)= [\theta(-(s+R\alpha))+\theta(s-R\alpha)]\cos\alpha+
    \theta(R\alpha-s)\theta(R\alpha+s)\cos{(\frac{s}{R})}
    \nonumber\\
    &&t_y(s)=
    -\mbox{sign}(s)[\theta(-(s+R\alpha))+\theta(s-R\alpha)]\sin\alpha-
    \theta(R\alpha-s)\theta(R\alpha+s)\sin{(\frac{s}{R})}\nonumber\\
    &&~~~~~~~~~~~~~~~~~~~~~~~~~~~~~~~~~~~~~~~~~~~~~~~~~~~~~~~~~~~~(\mbox{''open
    book''})
\end{eqnarray}
\begin{eqnarray}\label{omega}
&&t_x(s)=\theta(-s-\frac{\pi}{2}R)+\theta(s-\frac{\pi}{2}R)+\theta(\frac{\pi}{2}R-s)
\theta(\frac{\pi}{2}R+s)\cos{(\frac{s}{R})}
\nonumber\\
&&t_y(s)=\theta(\frac{\pi}{2}R-s)\theta(\frac{\pi}{2}R+s)\sin{(\frac{s}{R})}~~~~~~~~~~~~~~~~~~~(''\Omega-\mbox{curve}'')
\end{eqnarray}
The coordinate $s$ is counted from the center of the curves.

Despite simple problem formulation, the target setting has
pitfalls. In fact, absence of effective driving force apart from
the curved part of the wire results in vanishing of stationary
current in this section of a wire and, consequently, due to
continuity equation,   the current in any part of the wire. One
can make certain in that fact directly from the Eq. (\ref{kin}) by
integration with respect to the momentum.

We shall consider separately two cases, ballistic and kinetic
ones. By a ballistic wire we mean a wire shorter than the mean
free path, ends of which join the electron seas. In that case one
can neglect the scattering inside the wire and consider entering
electrons as equilibrium.

On the contrary, if mean free path is small as compared with the
wire length, the equilibrium is achieved inside the wire while
equilibrium states can differ in the right and the left sides of
the wire. This case needs to be solved by means of the kinetic
equation accounting for the collisions.

\subsection*{Ballistic wire}
In the ballistic limit we start from the Eq.(\ref{kin}) with
omitted right hand. Solving the kinetic equation in the second
order of electromagnetic field we find the correction to the
isotropic part of the distribution function. From the kinetic
equation it follows that the the isotropic part of the
distribution function should have different values on the ends of
the wire $s=\pm L/2$.  The difference of two limits $(\int dp
f(p)_{s= L/2}-\int dp f(p)_{s= -L/2})/\pi$ can be attributed to
the difference of concentrations on the contacts $\Delta n$, in
other words, to the difference of their chemical potentials
$\Delta\mu=\Delta n/(\partial n/\partial \mu).$  Taking into
account the electroneutrality reasons we should keep the
concentration on the wire ends, that can be done by adding a
static voltage such as $eV=\Delta\mu$.

The solution of linearized collisionless kinetic equation
satisfying the condition of equilibrium on the wire ends reads as
\begin{equation}\label{f1b}
   f_1^\omega =\exp{(\frac{i\omega s}{v})}\int_{\mp L/2}^s ds' \exp{(\frac{-i\omega
    s'}{v})}e{\cal E}^\omega(s')f_0', \end{equation}
    where   $f_0$ is the Fermi
function, prime means the derivation over energy
$\varepsilon=p^2/2m; ~~ L$ is the normalization length. The upper
~(lower) sign of the integral limit corresponds to $v> (<)~ 0.$

For $\Delta n $ we have
\begin{eqnarray}\label{dn}
    \Delta n=\frac{1}{\pi}\int_{-\infty}^{\infty} dp \bar{f}_2 =\frac{e}{2\pi m}\mbox{Re}
    \int_{-\infty}^\infty \frac{dp}{v^2}\int_{-\infty}^\infty ds {\cal E}^{-\omega}(s)f_1^\omega(s),
\end{eqnarray}
where $\bar{f}_2$ is the stationary part of the quadratic in
${\cal E}$ distribution function. Inserting expression (\ref{f1b})
into Eq.(\ref{dn}) one can obtain
\begin{equation}\label{dn1}
    \Delta n=-\frac{e^2}{2\pi m}\int_0^\infty
    \frac{dp}{v^2}f_0'\int_{-L/2}^{L/2}\int_{-L/2}^{L/2}ds
    ds'\sin{(\frac{i\omega (s-s')}{v})}
\mbox{Im}({\cal E}^{-\omega}(s){\cal E}^\omega(s')).\end{equation}
Taking into account symmetry of curves under consideration we
obtain for voltage $V$ from Eq.(\ref{dn1})
\begin{equation}\label{dn2}
    V=V_0\zeta \frac{\omega^2}{\pi (\partial n/\partial \mu)}\int_0^\infty
    \frac{dp}{v^2}f_0'\int_{-L/2}^{L/2}\int_{-L/2}^{L/2}ds
    ds'\sin{(\frac{i\omega (s-s')}{v})}
t_x(s)t_y(s').\end{equation} Here $$
V_0=\frac{e}{2m\omega^2}(E^\omega)^2.$$

 Introducing the space
Fourier transforms
$$\tilde{t}_{x,y}(q)=\int_{-L/2}^{L/2} ds~ t_{x,y}(s)e^{-iqs}$$
we arrive at:
\begin{equation}\label{Vcolless}
    V=V_0\zeta\frac{\omega^2}{\pi (\partial n/\partial \mu)}\int_0^\infty \frac{dp}{v^2}f_0'
    \mbox{Im}[\tilde{t}_x(\frac{\omega}{v})\tilde{t}_y(\frac{\omega}{v})]
.\end{equation} For degenerate Fermi gas for which ~ $\partial
n/\partial \mu=2/(\pi v_F)$ it follows from Eq.(\ref{Vcolless})
\begin{equation}\label{Vcolless1}
    V=-V_0\zeta\frac{\omega^2}{2v_F^2}
    \mbox{Im}[\tilde{t}_x(\frac{\omega}{v_F})\tilde{t}_y(\frac{\omega}{v_F})]
.\end{equation}
 We shall exemplify the general result (\ref{Vcolless}) by
means of the case of ''hard angle'' (Fig.1 a), where voltage takes
the form:
\begin{equation}\label{angle1}
V=V_0\zeta\frac{1}{\pi (\partial n/\partial
\mu)}\sin{(2\alpha)}\int_0^\infty dp f_0'
    [2\sin{(\eta/2)}-\sin{\eta}],
\end{equation}
$\eta=\omega L/v$. In particular case of degenerate Fermi system
Eq.(\ref{angle1}) leads to:
\begin{equation}\label{angle2}
V=-V_0\zeta\frac{1}{2}\sin{(2\alpha)}
    [2\sin{(\eta_F/2)}-\sin{\eta_F}], ~~~~~~~(\eta_F=\eta |_{v=v_F}).
\end{equation}
 The result (\ref{angle2}) stays limited when
$L\to\infty$, this implies that the voltage is formed by a finite
curved part of the wire. Nevertheless, the remaining parts $s\sim
L$ of the wire also participate in the voltage: the evidence of
that is presence of time-of-flight oscillations at frequencies
$\omega_N= 4N\pi v_F/L$ ($N$ is integer) in (\ref{angle2}) (which
evidently survive for any considered shape of wires, containing
straight parts). The value of voltage, averaged with respect to
oscillations vanishes in the limit of large $L$. Just this value
survives if the damping will be taken into account. The limited
nature of result for ''hard angle'' leads to finite contribution
of the curved part.

\subsection*{Kinetic approach} Here we solve the kinetic equation
(\ref{kin}) with collision term for infinitely long quantum wire,
in assumption that equilibrium distribution is established inside
the wire. The linear in ${\cal E}$ correction
$f_1(t)=\mbox{Re}(f_1^\omega e^{-i\omega t})$ to the equilibrium
distribution function obeys to the equation
\begin{equation}\label{f1}
    (-i\omega+iqv)\tilde{f}_1^{\omega}-e\tilde{\cal E}^{\omega}(q)vf_0'=
    -\nu[\tilde{f}_1^\omega-<\tilde{f}_1^\omega>].
\end{equation}
 The solution of Eq.(\ref{f1}) reads
\begin{equation}\label{f1a}
    \tilde{f}_1^{\omega}=-ie\tilde{\cal E}^\omega(q)vf_0'\frac{\omega+qv}{q^2v^2-\omega^2-i\omega\nu}
\end{equation}
For calculation of photocurrent we need in stationary quadratic in
${\bf E}$ part of distribution function. It satisfies the equation
\begin{equation}\label{f2}
    (\nu +iqv)\tilde{\bar{f}}_2(p;q)-\frac{e}{4}\sum_{q'}
    [\tilde{\cal E}^{-\omega}(q-q')\frac{\partial \tilde{f}_1^\omega(p;q')}{
    \partial p}+ (\omega \rightarrow -\omega)] =\nu <\tilde{\bar{f}}_2(p;q)>
\end{equation}
In an infinitely long wire the current tends to zero. This follows
from infinite resistance of the system and finiteness of the size
where the external field drags electrons. Hence the static drag
current is compensated by the static ohmic leakage current caused
by appearing static electric field ${\cal E}_0$:
$$\tilde{j}(q)+\tilde{\sigma}(q)\tilde{\cal E}_0(q)=0,$$ where
 $\tilde{\sigma}(q)$ is the linear conductivity. The
photovoltage is
\begin{equation}\label{V}
    V\equiv \lim_{q\rightarrow 0}\tilde{\cal E}_0(q)=-\frac{\lim_{q\rightarrow
0}\tilde{j}(q)}{\tilde{\sigma}(0)}.
\end{equation}
 Thus, the determination of photovoltage requires  finding the limit of
space Fourier-component of photocurrent $\tilde{j}(q) $ at $q
\rightarrow 0$.  Using Eqs.(\ref{f2}) and (\ref{f1a}) we obtain
from Eq.(\ref{V})
\begin{equation}\label{V1}
V=-\frac{e^3}{(2\pi)^2\tilde{\sigma}(0)}\int_{-\infty}^\infty dp
f_0'v^2\frac{\partial(\tau v)}{\partial p}\int_{-\infty}^\infty dq
q|\tilde{\cal
E}^{\omega}(q)|^2\frac{\omega\nu}{((qv)^2-\omega^2)^2+(\omega\nu)^2},
\end{equation}
where $$\tilde{\sigma}(0)=-\frac{e^2}{\pi}\int_{-\infty}^\infty dp
f_0'v^2\tau.$$ With taking into account integration over $q$ the
value $|\tilde{\cal E}^{\omega}(q)|^2$ in (\ref{V1}) can be
replaced by $2\mbox{Im}(E_x^\omega
E_y^{-\omega})\mbox{Im}[\tilde{t}_x(q)\tilde{t}_y^*(q)].$

At $\nu \to 0$  Eq.(\ref{V1}) is simplified
\begin{equation}\label{V2}
   V=-\frac{e^3}{4\pi\tilde{\sigma}(0)}\int_{-\infty}^\infty dp
f_0'v^2\frac{\partial(\tau v)}{\partial p}\int_{-\infty}^\infty dq
q|\tilde{\cal E}^{\omega}(q)|^2\delta((qv)^2-\omega^2).
\end{equation}
The physical meaning of the Eq.(\ref{V2}) is very simple: in the
presence of external alternating field the curvature  induces
package of waves  $|\tilde{\cal E}^{\omega}(q)|^2$ any of which
accelerates electrons moving with the velocity of the wave
$v=\omega/q$.

In the specific case of degenerate Fermi statistics Eq.(\ref{V2})
reads
\begin{equation}\label{V3}
   V=-V_0\zeta\beta\frac{\omega^2}{2v_F^2}
   \mbox{Im}[\tilde{t}_x(\frac{\omega}{v_F})\tilde{t}_y(\frac{\omega}{v_F})].
\end{equation}
where $\beta=1+2\partial(\ln\tau)/\partial(\ln\varepsilon_F)$. If
the relaxation time doesn't depend on electron energy
Eq.(\ref{V3}) reduces to the collisionless limit
Eq.(\ref{Vcolless1}). Though both considerations give voltage
independent on the scattering strength, more accurate approach
Eq.(\ref{V3})  depends on the character of scattering (via energy
dependence of $\tau$), despite smallness of the scattering rate.
In the case of weak scattering the mean free path exceeds the
length of curved domain and thus one should think the scattering
does not affect the voltage. The discrepancy  can be explained by
the fact that the voltage is formed on the same distance as the
conductivity, namely mean free path $v\tau$, or, in other words,
on the distance where the scattering occurs.

Let us consider another limit when mean free path is less than the
length of the curved part. Neglecting $q$ in the denominator of
the ratio in Eq.(\ref{V2}) one can get to the expression
\begin{equation}\label{V11}
V=-\frac{e^3}{2\pi\tilde{\sigma}(0)}\int_{-\infty}^\infty dp
f_0'v^2\frac{\tau}{\omega(\omega^2\tau^2+1)}\frac{\partial(\tau
v)}{\partial p}\int_{-\infty}^\infty ds k(s),
\end{equation}
where $k(s)=(t_y'(s)t_x(s)-t_y(s)t_x'(s))$ is the curvature. The
integral $\int ds k(s)$  equals to the total angle of rotation of
the vector  ${\bf t}$. This result physically follows from
locality of static field production in the case of small mean free
path: that means that the static field ${\cal E}(s)$ in the local
case can be determined by nothing else as local curvature $k$. The
universality of Eq. (\ref{V11}) gives immediately the same result
for the cases of ''hard angle'' and ''open book'' and zero result
for ''$\Omega$-curve'' (here we emphasize that all hard angles on
curves are supposed to be smoother  than $l$).

\subsection*{Specific shapes}

Expressions for induced voltage  when mean free path is comparable
than the  curved part can be obtained from Eq.(\ref{V1}) by
substitution of specific expressions for $\tilde{t}_{x,y}(q)$,
while Eqs.(\ref{V2}) and (\ref{V3}) refer to the limit of large
mean free path.

For the  ''hard angle''  $\tilde{\bf t}(q)$ has singular behavior
at $q=0$:
$\tilde{t}_x(q)=2\pi\delta(q)\cos\alpha,~~\tilde{t}_y(q)=(2i/q)\sin\alpha$.
This singularity  originates from behavior ${\bf t}(s)$ at
$s\to\pm\infty$. Substituting $\tilde{\bf t}(q)$ to Eq.(\ref{V1})
we find
\begin{equation}\label{angle3}
V=-V_0\zeta\beta\sin{(2\alpha)}\frac{\omega\tau}{1+(\omega\tau)^2}
\end{equation}

  In the case of ''open book'' (Fig.1 b) at $\nu\to 0$  and zero temperature the voltage
  the Eq. (\ref{V2})
  yields
\begin{eqnarray}
V=-V_0 \zeta\beta F(\xi),
\end{eqnarray} where

\begin{eqnarray} \label{opbook}
F(\xi) = \frac{2\xi-2\xi
\cos(2\alpha)\cos(2\alpha\xi)-(1+\xi^2)\sin(2\alpha)\sin(2\alpha\xi)}{2(1-\xi^2)^2},
\end{eqnarray}
$\pi -2\alpha$ is the angle at the vertex (see Fig.1 b),
$\xi=R\omega/v_F$, $R$ the radius of the circular part. If $\xi$
is small or large  (corresponding to small and large radius $R$),
$ F(\xi)\approx (1-\cos(2\alpha)-\alpha\sin(2\alpha))\xi $ at
$\xi\ll 1$ and
    $F(\xi)\approx -\sin(2\alpha)\sin(2\alpha\xi)/(2\xi^2)$ at $\xi\gg 1$.

    Eq.(\ref{V2}) and hence Eq. (\ref{opbook}) are valid in the
    extreme case of $\nu\to 0$. If $R\to 0$ the ''open book'' converts to the ''hard angle''
    and the
    Eq. (\ref{opbook})
    should convert to Eq.(\ref{angle3}) at $\omega\tau\gg 1$. This is not so.
    In fact here we have competition of two small parameters,
    $\xi$ and $1/\omega\tau$.
Taking into account singular contributions to
    the Fourier transform $\tilde{\bf
    t}(q)$ (the same as in the case of ''hard angle'')  give
    an additive contribution to the voltage
\begin{equation}\label{add}
    -V_0\zeta\beta\sin{(2\alpha)}\frac{1}{\omega\tau}
    \end{equation}
exactly coinciding with the ''hard angle'' result (\ref{angle3})
at $\omega\tau\gg 1$.

For $\Omega$-like wire the function $F(\xi)$ is replaced by
\begin{eqnarray}\label{omega1}
F(\xi)=\frac{2\xi^2}{(1-\xi^2)^2}\left[\xi
\cos^2{(\frac{\pi\xi}{2})}-\frac{1-\xi^2}{2}\sin{(\pi\xi)}\right].
\end{eqnarray}
Note, that in this case the contribution (\ref{add}) does not
appear, because straight parts of the curve are parallel. In
limiting cases of large and small $\xi$ this function behaves as $
F(\xi)\approx -(\pi-2)\xi^3$ at $\xi\ll 1$ and
    $F(\xi)\approx \sin\pi\xi+(1+\cos\pi\xi)/\xi$ at $\xi\gg 1$.

\subsection*{Discussion}
The considered ballistic and kinetic regimes differs by the
participation of scattering in the voltage appearance. In the
ballistic regime electrons enter the wire being in equilibrium
differing in the entrance and exit. In the kinetic regime the
scattering and equilibrium establishment  occurs inside the wire.
This has no affect on the order of the voltage magnitude being
mainly determined by the parameter $V_0$ in both cases.
 Qualitatively,  $eV_0$ is the mean kinetic energy which
electron obtains from the alternating field. The estimations give
$V_0\simeq 10^{-6}$ V for $E=1$V/cm, $\omega=10^{11}s^{-1}$, and
electron mass of GaAs $m=0.07\cdot 10^{-27}$g.

The ballistic voltage experiences time of flight oscillations.
They smear out by scattering. In a particular case of   ''hard
angle'' this results in disappearance of the voltage in the
kinetic regime. Under very strong scattering the static field
becomes local being determined by the local curvature; hence the
global voltage is determined by a global geometry of the wire,
namely total angle of rotation of the tangent. In the case of
moderate scattering the voltage contains two contributions, one of
which depends on the angle between straight entrance and exit and
other depends on the local geometry of the wire. The first of
these contributions vanishes if the angle is $\pi$ (for example,
in $\Omega$-curve).

Note, that in the kinetic regime the voltage goes to finite limit
if the length tends to infinity. This limit as a rule has the same
order of magnitude as in the ballistic regime, but, unlike the
latter, is determined by the energy dependence of the relaxation.
This difference is specified by difference of relaxation: inside
the wire in the kinetic case and outside the wire in the ballistic
case.

The considered here effect occupies its place in a rank of other
photoelectric effects, including photon drag
\cite{dan,grinb,gibs}, photogalvanic effect \cite{belin, ivch,
we}, quantum pumps \cite{tau, brouw, moskal, we1}, {\it etc}. It
is desirable to compare them with each other. Unlike the
photogalvanic effect which occurs in macroscopically uniform
media, this curvature-induced effect has local character. When the
system size goes to infinity the produced voltage tends to the
fixed limit instead of growth.

Let there are multiple kinks on the wire distributed with constant
density. Then any kink will produce a fixed voltage, so that they
additively produce mean electric field (and corresponding current
in a short circuite regime). This variant reminds the photocurrent
in a spiral quantum wire \cite{spiral}. Both effects have  the
same physical origin: the photon drag due to the curvature-induced
effective momentum of the wave. The found voltage is independent
on the relaxation rate in the limit of weak relaxation (except for
the case of ''hard angle'', where the absence of relaxation leads
to the absence of voltage). At the same time the current in the
spiral grows in the limit of weak relaxation under
 resonant conditions
(when the velocity of wave coincides with the Fermi velocity).
This difference between effects results from the fact that the
kink produces wave package instead of a single wave in the case of
spiral curve  and averaging on momentums washes out the resonance
and the dependence on the collision rate.

The system under consideration  is akin  to quantum pumps by
locality of the alternating perturbation, but unlike them the
found (with relaxation taken into account) current disappears in
infinite system.

The asymmetry of considered systems is artificial. From this
viewpoint these systems are similar to the artificial antidot
lattices \cite{semidisk1,we1,kvon}.  Being planar, the curved
wires can be simply realized experimentally, in particular, in the
same antidot lattices by depleting conditions. Hence, there is
hope for a quick experimental verification.

\subsection*{Acknowledgments} The work was supported by grants of RFBR
(08-02-00506 and 08-02-00152) and Program of
SB RAS.

\end{document}